# Controlling spatial inhomogeneity in prototypical multiphase microstructures[*]

D. Frączek[1], R. Piasecki[2], W. Olchawa[2], R. Wiśniowski[2]

[1]Department of Materials Physics, Opole University of Technology,
Katowicka 48, 45-061 Opole, Poland
[2]Institute of Physics, University of Opole, Oleska 48, 45-052 Opole, Poland

A wide variety of real random composites can be studied by means of prototypes of multiphase microstructures with a controllable spatial inhomogeneity. To create them, we propose a versatile model of randomly overlapping super-spheres of a given radius and deformed in their shape by the parameter $p$. With the help of the so-called decomposable entropic measure, a clear dependence of the phase inhomogeneity degree on the values of the parameter $p$ is found. Thus, a leading trend in changes of the phase inhomogeneity can be forecast. It makes searching for possible structure/property relations easier. For the chosen values of $p$, examples of two and three-phase prototypical microstructures show how the phase inhomogeneity degree evolves at different length scales. The approach can also be applied to preparing the optimal starting configurations in reconstructing real materials.

## 1. Introduction

Materials composed of randomly distributed components of distinct phases appear both in nature and in synthetic products. Effective properties of such disordered heterogeneous materials can be connected with the physical properties of the relevant phases [1-3]. Some of them, *e.g.* an effective electrical or heat conductivity, are sensitive to the average spatial arrangement of the material components. In this context, it should be noticed that the methods of quantitative characterization (in a wide range) of the spatial morphology are significant ones in computational materials science. On the other hand, microstructure modelling makes prediction of effective properties of heterogeneous media easier. Thus, an efficient method for statistical three-dimensional (3D) reconstructing is an indispensable part of developing a reasonable model microstructure [4–17].

The phase spatial inhomogeneity belongs to the simplest statistical features of multiphase random systems. However, this quantity is length scale dependent. There are various approaches used to develop such measures. Mathematicians prefer to consider systems of points randomly placed. However, the more realistic models make use of digital information contained, *e.g.* in cross-sample micrographs or tomographic data. Thus, it is

---

worth to make one observation. The appropriate and reliable measures should be defined for random systems composed of *finite-sized* objects, here, unit pixels $1 \times 1$ for two-dimensional (2D) case or unit voxels $1 \times 1 \times 1$ for 3D.

To obtain *quantitative* characteristics for an average degree of spatial inhomogeneity over different length scales, the multiscale entropic descriptors, the $S_\Delta$ for binary and the $G_\Delta$ for grey level patterns have been developed [18, 19]. The measures were successfully applied in computational statistical physics, *e.g.* [20–22]. We recommend the short introduction related to the above entropic descriptors and 2D numerical example presented in Appendix of Ref. [19], and for 3D case in Appendix of Ref. [22]. In turn, when some knowledge about spatial arrangement of the *i*th *phase component* is needed, then a more detailed approach is necessary [23].

## 2. The phase entropic descriptors

Quite recently, the extended multiphase entropic descriptor $S_\Delta$ has been decomposed into phase-separated descriptors $S_{i,\Delta}$, $i = 1, 2, …, w$, which were denoted as $f_{i,\Delta}$ in Ref. [23]. The phase descriptor per cell for a multiphase material build of $w$ phases is defined by the formula

$$S_\Delta(k) = \sum_{i=1,...,w}(f_{i,\max} - f_i)/\lambda = \sum_{i=1,...,w} f_{i,\Delta}(k) \equiv \sum_{i=1,...,w} S_{i,\Delta}(k), \quad (1)$$

where $f_i = k_B \ln \Omega_i \equiv S_i$ denotes the Boltzmann entropy and $f_{i,\max} = k_B \ln \Omega_{i,\max} \equiv S_{i,\max}$ means its maximal theoretical value. In what follows, we set $k_B = 1$. The $\Omega_i(k)$ is the number of realizations for a "non-equilibrium" actual macrostate (AM) defined as a set $\{m_i(\alpha, k)\}$ of occupation numbers for overlapping sampling $\lambda$-cells of size $k \times k \times k$ in voxels, $\alpha = 1, 2, …, \lambda(k)$. Similarly, $\Omega_{i,\max}(k)$ describes the number of realizations for the "equilibrium" reference macrostate (RM) that relates to a maximally uniform configuration at a given discrete length scale $k$. The sum of $S_{i,\Delta}$ over the phases equals exactly the overall $S_\Delta$. The highest of the local maximums of $S_{i,\Delta}(k)$ quantifies a maximal spatial inhomogeneity per cell for the *i*th phase reached at the characteristic length scale $k_{\max}(\#)$. In turn, the first maximum, *i.e.*, that observed at scale $k_{\max}(1) \leq … \leq k_{\max}(\#) \leq …$, indicates formation of the *i*th phase clusters of a characteristic range of sizes, which are comparable with $k_{\max}(1)$. For further details, the reader is referred to Ref. [23].

## 3. Prototypical multiphase microstructures

Recently, an interesting model of super-spheres with versatile shapes has been considered for heterogeneous materials [24, 25]. In general, a $d$-dimensional super-sphere with radius $R$ is defined by

$$|x_1|^{2p} + |x_2|^{2p} + ... + |x_d|^{2p} \leq R^{2p}, \qquad (2)$$

where $x_i$ are Cartesian coordinates, $i = 1, ..., d$, and $p \geq 0$ is the deformation parameter indicating to what extent the particle shape has been deformed from that of a $d$-dimensional sphere ($p = 1$). The shape deformation parameter $p \geq 0$ allows changing the shape from convexity to concavity as $p$ passes downward through 0.5; *cf.* Fig. 1 in Ref. [25]. Here, we present such a shape evolution for the specified values of $p$.

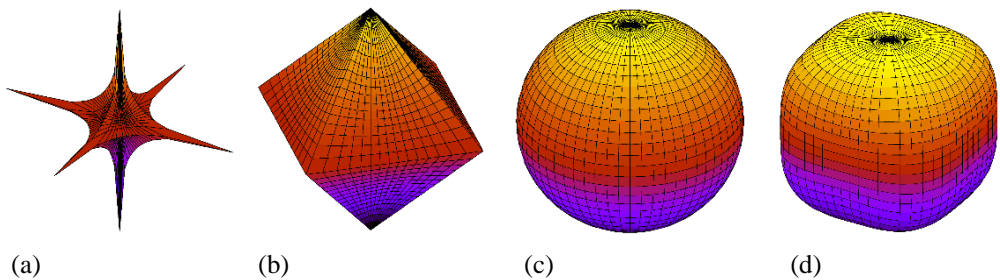

(a)      (b)      (c)      (d)

**Fig. 1.** Super-spheres with the chosen values of the shape deformation parameter $p$, which indicates to what extent the particle shape has been deformed from that of a three-dimensional sphere. (a) $p = 0.2$. (b) $p = 0.5$. (c) $p = 1$. (d) $p = 1.6$.

Recently, the simple model of overlapping spheres (with $p = 1$, using the present notation) of a radius specified from the two-exponent power-law (TEPL) has been applied to generate low-cost preferred configurations [22]. Here, we apply the same model using super-spheres given by (2). The centres of super-spheres are randomly distributed on a regular lattice. Since in our approach the super-spheres of a fixed radius are free to overlap, clusters of various sizes, shapes and volumes are created. It is worth noticing that the super-spheres are aligned to the underlying lattice. Thus, for $p \neq 1$, the super-spheres all possess the same orientation. According to our (not published) findings for 2D binary patterns, the random orientations of identical finite-sized objects like filled black ellipses can be reduced to those aligned to the lattice axes without impacting significantly the values of entropic descriptor. However, when some of the random orientations are accompanied with random translations of the underlying objects, the situation becomes more intricate. We would like to point out that the entropy-based spatial inhomogeneity measure is sensitive mainly to changes in arrangement of finite-sized objects. Nevertheless, this point is worth taking into account in a future project. Now, making use of shape deformed parameter $p$, we obtain a powerful tool for creating, in a controllable way, prototypes of random multi-phase microstructures. Such a model extends the recent approach discussed in Ref. [26]. However, an overlapping sphere model used therein was defined in a different context.

In this manner, complex inhomogeneous microstructures dependent on the value of parameter $p$ can effortlessly be obtained. This is a reason why we consider the impact of shape deformation parameter $p$ on evolution of the system's spatial inhomogeneity. In the present short paper, the two-phase microstructures (matrix plus inclusions) were created by black phase interpenetrating super-spheres randomly distributed on a regular lattice of size $128 \times 128 \times 128$. A few selected values of the parameter $p = 0.2, 0.5, 1$ and $1.6$ were employed. All the "non-equilibrium" configurations were generated for a fixed radius $R = 10.5$, equal concentrations of the black and white phases, and with the same random seed.

Now, as regards the model microstructures, the $p$-dependent, thus controllable evolution of (a) overall spatial inhomogeneity, *i.e.*, for both phases $S_\Delta(k;p)$, (b) for black phase $S_{\text{black},\Delta}(k;p)$, and (c) for white matrix phase $S_{\text{white},\Delta}(k;p)$ can easily be investigated.

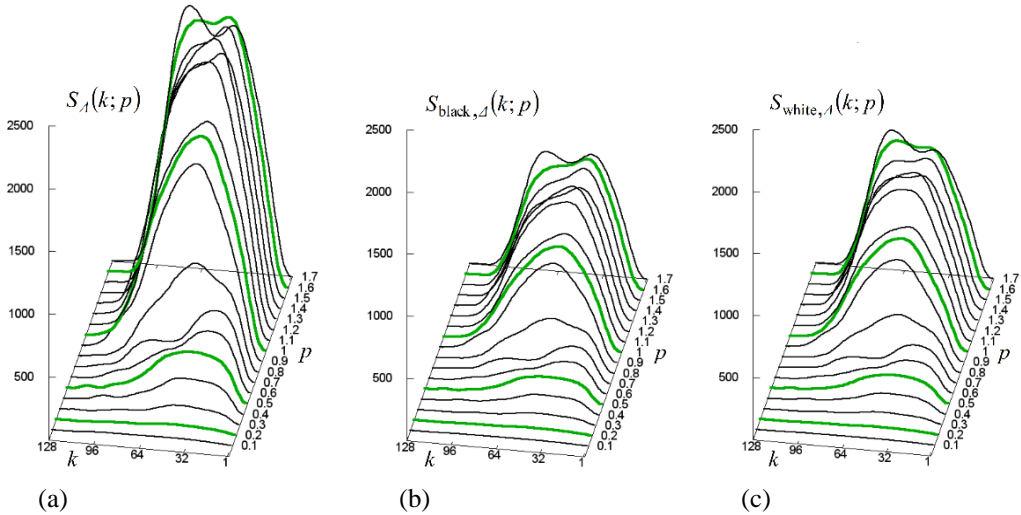

(a)          (b)          (c)

**Fig. 2.** The evolution of the two-phase system's spatial inhomogeneity measure as a function of length scale $k$ for fixed values of the deformation parameter $p = 0.1, 0.2, …, 1.7$. (a) The overall entropic descriptor $S_\Delta(k;p)$. (b) The black phase entropic descriptor $S_{\text{black},\Delta}(k;p)$. (c) The white phase entropic descriptor $S_{\text{white},\Delta}(k;p)$. The grey/green bold lines for the considered measure correspond to the selected values of the parameter $p = 0.2, 0.5, 1$ and $1.6$.

In each case, Figs. 2 (a), (b) and (c) show non-monotonic significant changes in the spatial inhomogeneity measures against the length scale $k$ for fixed values of the deformation parameter $p = 0.1, 0.2, …, 1.7$. A clear dependence of the phase inhomogeneity degree on the values of the parameter *p* is found. In general, the smaller the parameter *p* is, the lower the spatial inhomogeneity appears. As expected, at the same time, one can observe a shift of the first peak toward smaller length scales. Thus, a leading trend in changes in the phase inhomogeneity can be forecast, which makes searching for possible structure/property relationships easier. Additionally, for the selected *p*-cubes with the considered four values of the shape deformation parameter *p* (see the grey/green bold lines in Fig. 2), the corresponding cross-sections are illustrated in Figs. 3 (a), (b), (c) and (d).

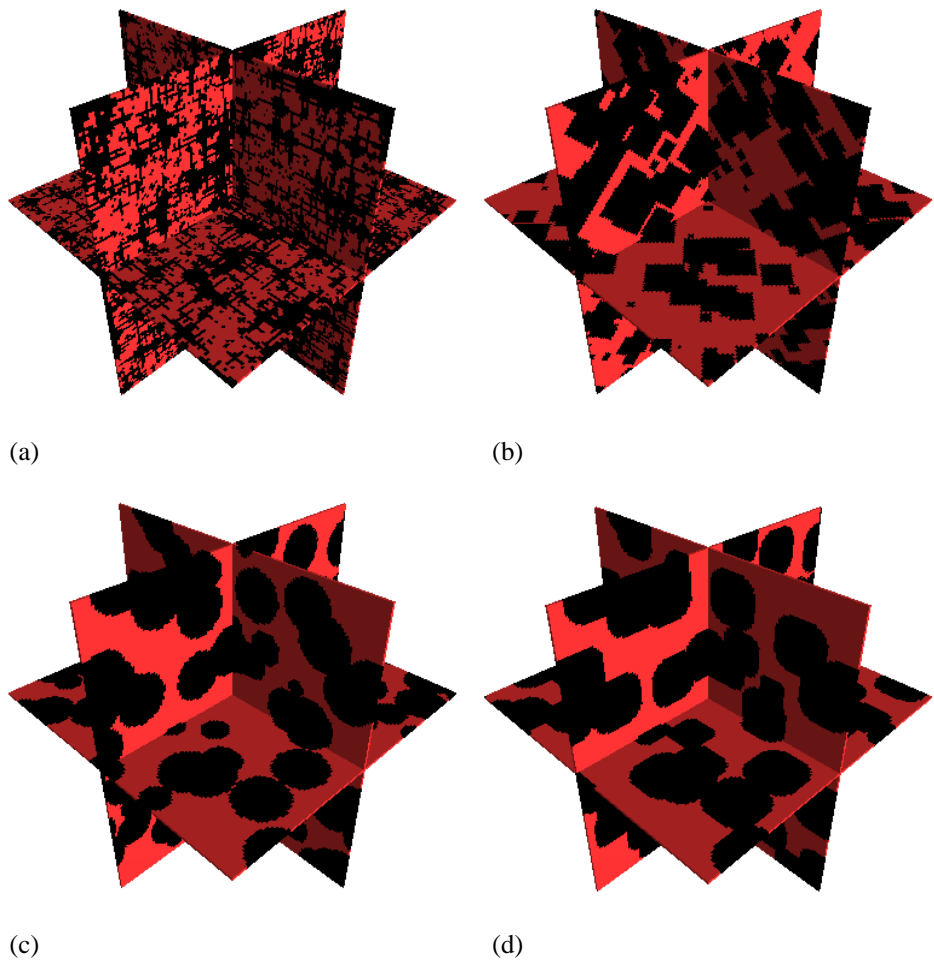

(a) (b)

(c) (d)

**Fig. 3.** The illustrative cross sections of the prototypical two-phase microstructures of size $128 \times 128 \times 128$ in voxels, for selected values of the parameter *p*. (a) $p = 0.2$. (b) $p = 0.5$. (c) $p = 1$. (d) $p = 1.6$. (In Fig. 2, the grey/green bold lines for the considered measure correspond to these prototypes.)

Additionally, in Fig. 4 (a), the phase entropic descriptors $S_{\text{black},\Delta}(k;p)$, $S_{\text{grey},\Delta}(k;p)$ and $S_{\text{white},\Delta}(k;p)$ corresponding to the black, grey and white phases of equal concentrations, respectively, are depicted as functions of length scale *k* and for the chosen two near values of shape deformation parameter, $p = 0.30$ and $0.45$. Notice that a much more complicated behaviour of the phase entropic descriptors can be found for the corresponding prototypes of the three-phase microstructure. This time, one can observe the irregular interplaying of the phase entropic descriptors in Fig. 4 (a). Such a behavior of the $S_{\text{black},\Delta}(k;p)$, $S_{\text{grey},\Delta}(k;p)$ and $S_{\text{white},\Delta}(k;p)$ is hard to predict. For completeness, an exterior view of the related $p$-cubes is shown in Figs. 4 (b) and (c), respectively. The complex case of three-phase microstructures needs further investigations.

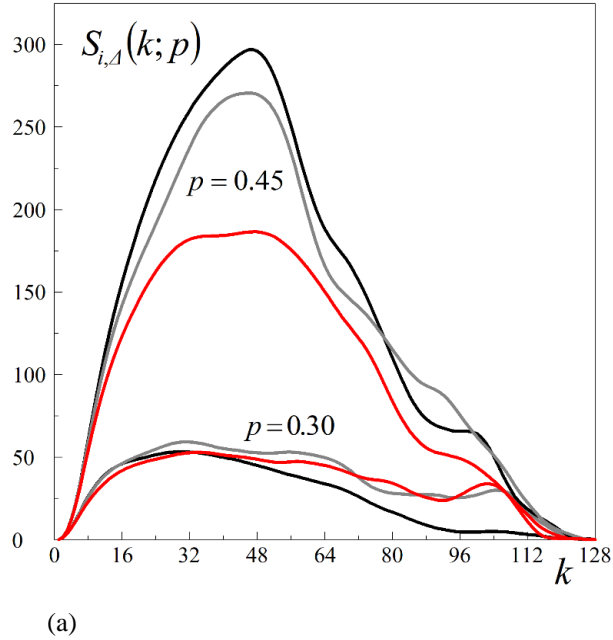

(a)

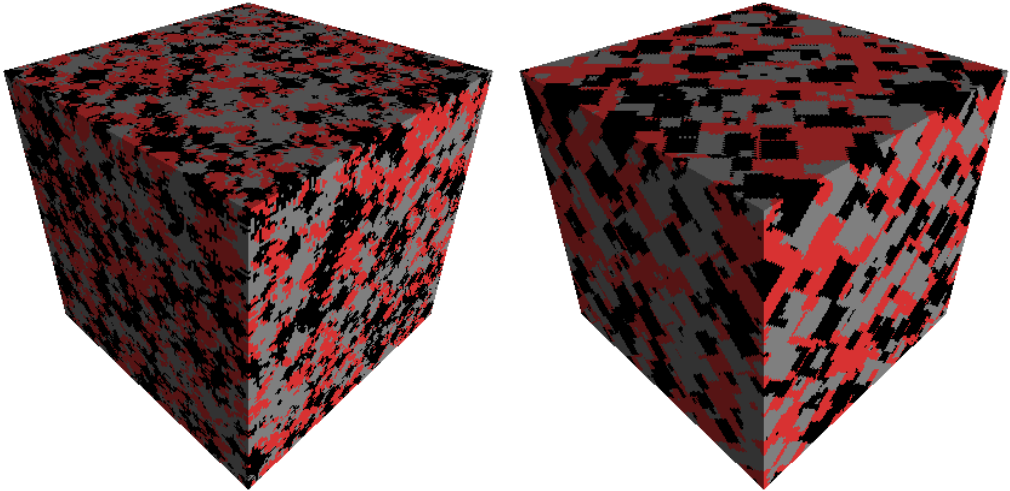

(b)　　　　　　　　　　(c)

**Fig. 4.** (a) The black phase entropic descriptor $S_{\text{black},\Delta}(k;p)$, the black line, the grey phase entropic descriptor $S_{\text{grey},\Delta}(k;p)$, the grey line, and the white phase entropic descriptor $S_{\text{white},\Delta}(k;p)$, the thick black/red line, for a three-phase system as a function of length scale $k$ for selected values of $p = 0.30$ (the bottom three lines) and 0.45 (the upper ones). (b) The three-dimensional exterior view of the related cube with $p = 0.30$. (c) The same for $p = 0.45$.

Finally, it should be noticed that three-phase microstructures corresponding, in the present notation, to the case $p = 1$ can also be found in the earlier mentioned Ref. [26]. However, the overlapping sphere model used therein was defined for different purposes, namely for building of percolation phase diagrams.

## 4. Concluding remarks

Many types of real random composites can be analysed using prototypes of the multiphase microstructures with a controllable spatial inhomogeneity. In order to create them, a low-cost model of randomly overlapping super-spheres can be applied. In turn, the versatile entropy method of quantitative characterization of spatial features of any phase microstructure can efficiently be used in searching for possible structure/property relationships. For instance, the expected dependence of an effective DC conductivity on the spatial inhomogeneity of phases of a multiphase complex composite becomes a much easier task. Using a simple variant of the real-space renormalization group approach [27], the related investigations were done (DF&RP) with promising results for a two-phase microstructure.